# Semiconductor quantum well magnetic memory using confinement from proximity exchange fields for high magnetoresistances in a field-effect transistor

William S. Rogers[1,2], and Jean Anne Incorvia[2]

[1]*Graduate Program in Applied Physics, Northwestern University, Evanston, IL, USA*
[2]*Chandra Department of Electrical and Computer Engineering, The University of Texas at Austin, Austin, TX, USA*
Email: *williamrogers2029@u.northwestern.edu*

**Abstract**
There is a growing demand for highly-performant memories and memristive technologies for use in in-memory computing. Magnetic tunnel junctions (MTJs) have thus far addressed this need in the field of spintronics. Despite their low write power and high speeds, MTJs are limited by their modest on/off ratio at room temperature, which motivates a search for beyond-MTJ spintronic devices. In this work, we propose a device that uses two layers of ferromagnetic insulator (FMI) cladding a semiconductor QW, which is able to modulate the QW bandgap via electronic confinement resulting from proximity magnetization at the interfaces of the quantum well depending on the relative magnetization of the FMI layers. We predict that this device has the potential for very high magnetoresistances (MRs) possibly exceeding 10,000% at room temperature. We also predict that this device will operate with maximal MR in charge neutrality, and that electrostatic gating may promote the device to act as a magnetic memtransistor. This motivates the search for candidate materials and ultimately experimental demonstration of magnetic QW memories or memtransistors, which may have the potential to advance the state of the art in logic, memory, or neuromorphic circuits.

**Introduction**
Spin-based devices have long held promise for new memory and computing applications, such as magnetic random-access memory (MRAM) for embedded applications [1], tunnel magnetoresistance (TMR)-based magnetic sensors for improved hard drive densities, and domain wall racetracks [2] with use in memory and neuromorphic computing[3]. The magnetic tunnel junction (MTJ) has been by far the most applicative nanodevice to come out of this effort to-date, with sub-ns switching times [4], low power consumption for highly efficient information write, and modest on/off ratios (less than $10^3$%) for magnetic information readout. This is not without drawbacks: there is a tradeoff between on-state conductivity and TMR via modifying the tunnel barrier thickness [5], high-TMR MTJs are challenging to grow (such as with pinholes in sputter-grown devices [6]), or at scale, and other memories such as conductive bridge RAM (CBRAM), phase change memory (PCM), ferroelectric random-access memory (FeRAM) or NAND flash outperform the MTJ on/off ratio by multiple orders of magnitude [7]. Ultimately, MTJ devices which rely on the MgO $\Delta_1$ symmetry filtering and finite spin polarization of the CoFe $\Delta_1$ band are thus far limited to a < 700% TMR at room temperature and rapidly decrease in TMR upon symmetry breaking from barrier defects [5].

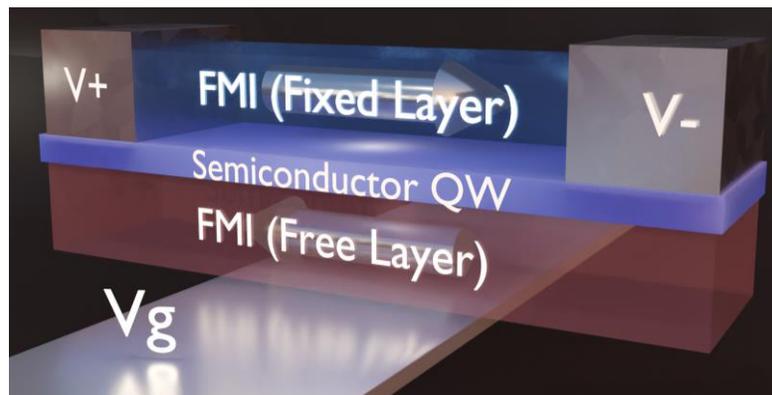

**Fig. 1** Schematic of the proposed device with antiparallel (AP) magnetic configuration.

In Fig. 1, we show a schematic of our proposed device which relies on magnetic readout via modulation of the conduction band edge in a proximity-magnetized semiconductor quantum well (MQW) sandwiched between two ferromagnetic insulators (FMI). As shown in Fig. 2, depending on the alignment of the two FMI interfaces, parallel

(P) or antiparallel (AP), the exchange fields will constructively or destructively add, thus modulating the spin-confinement of the electrons within the QW and modulating the conduction band edge. This will give rise to a novel magnetoresistance arising from magnetic exchange confinement of the electrons and bandgap modulation. Then, a readout of the transverse conductivity along the QW (across V+ and V- in the Fig. 1 schematic) will probe the 0 or 1 state of the device. Modulation of a gate voltage $V_g$ may additionally allow for operation as a magnetic memtransistor, and could double as a spin-orbit torque write line to inject spin currents into the FMI for magnetic write capability [8].

Here, we employ a model Hamiltonian to show that MQW devices have the potential to facilitate very high MRs, possibly up to or beyond room temperature, in a simple device structure without need for perfect electrode spin polarization or complex k-dependent electronic phenomena. The device only relies on the FMI to induce a modest proximity-based exchange splitting in the QW interface, which we believe may be more robust than the MRs based on spin transport. First, we derive an effective mass tight-binding model, which we use to semiclassically evaluate the channel sheet conductivity of field-gated MQW thin films to inform the design of future MQW field-effect transistors (MQW-FETs).

**Effective mass model:**
We will consider the low-energy effective mass model of the Schrödinger equation to model an n-type semiconductor with electron effective mass $m^*$, with Rashba-type spin orbit coupling parameterized by $\eta$, conduction band edge $E_c$, and a proximity-induced magnetic exchange splitting field $\boldsymbol{m}(\boldsymbol{R}) \cdot \boldsymbol{\sigma}$ as a function of space $\boldsymbol{R}$, where $\boldsymbol{\sigma}$ is the vector of Pauli matrices acting on the electron spin degree of freedom, and $\sigma_0$ is the identity matrix in spin space:

$$\hat{H} = \left(-\frac{\hbar^2}{2m_e^*}\boldsymbol{\nabla}^2 + E_c\right) \otimes \sigma_0 + i\eta(\boldsymbol{\nabla} \times \boldsymbol{\sigma}) \cdot \hat{z} + \boldsymbol{m}(\boldsymbol{R}) \cdot \boldsymbol{\sigma} \quad (eq.1)$$

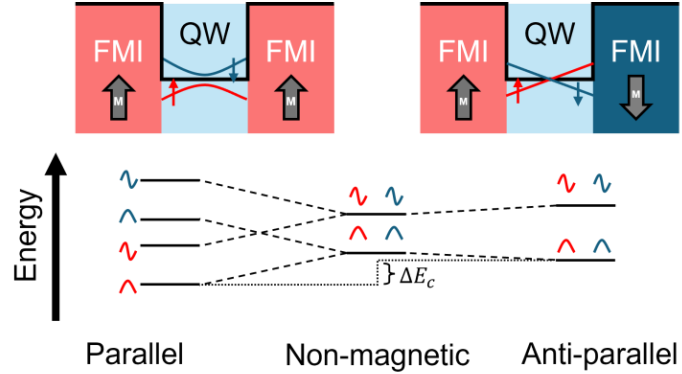

**Fig 2.** The conduction band edge of the MQW-FET device is pictured under differing magnetic configurations. The parallel, or low-resistance state of the magnetic QW device is shown at left. The spin degeneracy of each non-magnetic QW subband is lifted in the P case, while the states are spin-degenerate in the AP case, shown at right.

While perhaps important for a system with heavier elements in an ultrathin QW, we will neglect the spin-orbit interaction in this work ($\eta = 0$), assuming that the magnetic exchange is much greater, and construct a minimal 4x4 tight-binding model which captures the essential phenomena regarding the four lowest-energy states in the QW with translation symmetry in x and y. The device is broken into two sites in z spaced apart by $a = \Delta z/2$, where $\Delta z$ is the total QW thickness; this constitutes the pseudospin degree of freedom over the left (L) and right (R) site, and $\tau_i$ are the pauli matrices in pseudospin space. We fix the magnetization field in z such that $\boldsymbol{m}(\boldsymbol{R}) \rightarrow Jm_{L,R}\hat{z}$, with $m_{L,R} \in \pm 1$ and an average exchange splitting J over each site. The tight-binding parameter between the two sites is given as $t = \frac{\hbar^2}{2m_e^* a^2} = \frac{2\hbar^2}{m_e^* \Delta z^2}$ after Taylor approximation of the parabolic dispersion, with a corresponding 2t added to the diagonal of the Hamiltonian:

$$H(k_x, k_y) = \left(\frac{\hbar^2(k_x^2 + k_y^2)}{2m_e^*} + E_c + 2t\right)\tau_0 \otimes \sigma_0 - t\tau_x \otimes \sigma_0 + Jm_L\tau_+\tau_- \otimes \sigma_z + Jm_R\tau_-\tau_+ \otimes \sigma_z \quad (eq.\,2)$$

Or for clarity, defining $\frac{\hbar^2(k_x^2+k_y^2)}{2m_e^*} + E_c = E_0$

$$\hat{H} = \begin{bmatrix} E_0 + Jm_L + 2t & 0 & -t & 0 \\ 0 & E_0 - Jm_L + 2t & 0 & -t \\ -t & 0 & E_0 + Jm_R + 2t & 0 \\ 0 & -t & 0 & E_0 - Jm_R + 2t \end{bmatrix} \quad (eq.\,3)$$

Using this, we can exactly determine the energy spectrum for the P and AP memory states. The energy diagram for the lowest QW subband is shown in Fig 2:

$$E_P \in \{-J + t + E_0, -J + 3t + E_0, J + t + E_0, J + 3t + E_0\}; \quad (eq.\,4)$$

$$E_{AP} \in \left\{-\sqrt{J^2 + t^2} + 2t + E_0, -\sqrt{J^2 + t^2} + 2t + E_0, \sqrt{J^2 + t^2} + 2t + E_0, \sqrt{J^2 + t^2} + 2t + E_0\right\} \quad (eq.\,5)$$

Thus, we can calculate the modulation of the conduction band from the P to AP state, defining $\Delta E_c = E_{c,AP} - E_{c,P}|$:

$$\Delta E_c = J + t - \sqrt{J^2 + t^2} \quad (eq.\,6)$$

$$= J + \frac{2\hbar^2}{m_e^*\Delta z^2} - \sqrt{J^2 + \frac{4\hbar^4}{m_e^{*2}\Delta z^4}} \quad (eq.\,7)$$

For a material with an effective interfacial exchange splitting of $J = 0.2$ eV, electron quasiparticle mass $m_e^* = 0.18\,m_0$ (to mimic an optimistic, highly engineered GdN-based heterostructure [9]), and an ultrathin QW with thickness $z = 2$ nm, this results in a large conduction band edge modulation by approximately 120 meV before screening effects will renormalize the conduction band offset to maintain charge neutrality in the device. This presents a challenge to readout, as (in comparison) electrostatic gating will instead modulate the Fermi level and allow for a large swing in carrier concentration. One path to mitigate the screening while still sensing an $E_c$ modulation may entail the design of a heterostructure utilizing band bending, however we believe that a modest hole population in an intrinsic semiconductor may be another promising path to ensuring charge neutrality in steady state; the device essentially will operate by a dynamic modulation of the bandgap. Thus, we also should consider the modulation of the valence band edge $\Delta E_v = E_{v,AP} - E_{v,P}$, where $m_h^*$ is the (negative) hole effective mass and $J_h$ is the hole exchange splitting ($m_e^*, J_e$ now referring to the electron bands):

$$\Delta E_v = -J_h + \frac{2\hbar^2}{m_h^*\Delta z^2} + \sqrt{J_h^2 + \frac{4\hbar^4}{m_h^{*2}\Delta z^4}} \quad (eq.\,8)$$

For a system with an intrinsic bulk gap $E_g^0$, QW gap $E_g^C$, P bandgap $E_g^P$, and AP bandgap $E_g^{AP}$, we should expect:

$$E_g^C = E_g^0 + t_e - t_h = E_g^0 + \frac{2\hbar^2}{\Delta z^2}\left(\frac{1}{m_e^*} - \frac{1}{m_h^*}\right); \quad (eq.\,9)$$

$$E_g^P = E_g^0 + \frac{2\hbar^2}{\Delta z^2}\left(\frac{1}{m_e^*} - \frac{1}{m_h^*}\right) - J_e - J_h; \quad (eq.\,10)$$

$$E_g^{AP} = E_g^0 + \frac{4\hbar^2}{\Delta z^2}\left(\frac{1}{m_e^*} - \frac{1}{m_h^*}\right) - \sqrt{J_e^2 + \frac{4\hbar^4}{m_e^{*2}\Delta z^4}} - \sqrt{J_h^2 + \frac{4\hbar^4}{m_h^{*2}\Delta z^4}} \quad (eq.\,11)$$

Thus the modulation in bandgap from the P to AP state, $\Delta E_g$, is given as

$$\Delta E_g = J_e + J_h + \frac{2\hbar^2}{\Delta z^2}\left(\frac{1}{m_e^*} - \frac{1}{m_h^*}\right) - \sqrt{J_e^2 + \frac{4\hbar^4}{m_e^{*2}\Delta z^4}} - \sqrt{J_h^2 + \frac{4\hbar^4}{m_h^{*2}\Delta z^4}} \quad (eq.\,12)$$

With a corresponding density of states DOS $D(E)$ near the band edges, which we have shown for a QW below [10]. We define $d_{v,c}$ as the intrinsic valence and conduction band valley degeneracy, and note the single spin in the band edges in the P case. $\Theta$ is the Heaviside step function. We set $E = 0$ to the middle of the bandgap.

$$D_P(E) = d_v \frac{|m_h^*|}{2\pi\hbar^2}\Theta(-E - E_g^P/2) + d_c \frac{|m_e^*|}{2\pi\hbar^2}\Theta(E - E_g^P/2) \quad (eq.\,13)$$

$$D_{AP}(E) = d_v \frac{|m_h^*|}{\pi\hbar^2}\Theta(-E - E_g^{AP}/2) + d_c \frac{|m_e^*|}{\pi\hbar^2}\Theta(E - E_g^{AP}/2) \quad (eq.\,14)$$

Then, the carrier concentration $n, p$ can be calculated as a function of Fermi level $E_f$, approximating the Fermi-Dirac function using Boltzmann statistics:

$$n_P(E_f) = \frac{d_c|m_e^*|k_BT}{2\pi\hbar^2}e^{-\frac{E_g^P/2 - E_f}{k_BT}}; \quad p_P(E_f) = \frac{d_v|m_h^*|k_BT}{2\pi\hbar^2}e^{-\frac{E_f + E_g^P/2}{k_BT}};$$

$$n_{AP}(E_f) = \frac{d_c|m_e^*|k_BT}{\pi\hbar^2}e^{-\frac{E_g^{AP}/2 - E_f}{k_BT}}; \quad p_{AP}(E_f) = \frac{d_v|m_h^*|k_BT}{\pi\hbar^2}e^{-\frac{E_f + E_g^{AP}/2}{k_BT}}$$

$$(eq.\,15 - 18)$$

So that the intrinsic carrier concentration in equilibrium $n_i = \sqrt{np}$, is given by

$$n_{i,P} = \frac{k_BT}{2\pi\hbar^2}\sqrt{|d_c d_v m_e^* m_h^*|}\, e^{-\frac{E_g^P}{2k_BT}} \quad (eq.\,19)$$

$$n_{i,AP} = \frac{k_BT}{\pi\hbar^2}\sqrt{|d_c d_v m_e^* m_h^*|}\, e^{-\frac{E_g^{AP}}{2k_BT}}. \quad (eq.\,20)$$

We can calculate the sheet conductivity in each case, and thus the on/off ratio in intrinsic operation. We will assume the electron and hole mobility are approximately constant, defining the conductivity $\sigma$ and magnetoresistance ratio

$$\sigma = e(\mu_e n + \mu_h p); \quad (eq.\,21)$$

$$\text{MR} = \frac{\sigma_P - \sigma_{AP}}{\sigma_{AP}} = \frac{(\mu_e + \mu_h)n_{iP}}{(\mu_e + \mu_h)n_{i,AP}} - 1 = \frac{1}{2}e^{\frac{\Delta E_g}{2k_BT}} - 1$$

$$= \frac{1}{2}\exp\left[\frac{1}{2k_BT}\left(\frac{2\hbar^2}{\Delta z^2}\left(\frac{1}{m_e^*} - \frac{1}{m_h^*}\right) + J_e + J_h - \sqrt{J_e^2 + \frac{4\hbar^4}{m_e^{*2}\Delta z^4}} - \sqrt{J_h^2 + \frac{4\hbar^4}{m_h^{*2}\Delta z^4}}\right)\right] - 1. \quad (eq.\,22)$$

It is of particular interest that the intrinsic gap $E_g^0$ drops out of the final expression for the MR, which may permit a wide range of semiconducting materials in these devices, if one tolerates an arbitrarily large P-state resistance. We must also note the caveat that this expression assumes $J \gg k_BT$, which implies that the spin-split band edges in the P case will dominate transport. Experiments at low $\Delta E_g$ will outperform the above expression, as our model excludes higher sub-bands, and this may be qualitatively fit with $\text{MR} = \cosh\left(\frac{\Delta E_g}{2k_BT}\right) - 1$ to correct for the resulting near-degeneracy in spin at low $J$ in the P case. One can extend this to consider the behavior of a MQW film that is uniformly back-gated, thus promoting the device to act as a magnetic memtransistor. We will assume a charge density $\frac{Q}{A} = \frac{V_g \epsilon_0 \epsilon_r}{t}$, for a YIG gate dielectric with $\epsilon_r = 12$ [11]; and thickness t = 2.5 nm. From [10], we define

$$n = \frac{1}{2}\left[\frac{Q}{e} + \sqrt{\left(\frac{Q}{e}\right)^2 + 4n_i^2}\right], p = \frac{1}{2}\left[-\frac{Q}{e} + \sqrt{\left(\frac{Q}{e}\right)^2 + 4n_i^2}\right], \quad (eq.\,23, 24)$$

for devices near equilibrium, which can be combined with Eq. 22. We also assume that scattering deriving from film roughness, grain boundary, and impurity scattering dominate in all temperature regimes, which may be characteristic of initial experiments, and fix the scattering lifetime $\tau_e = \tau_h = 10$ fs, to express $\mu_{e,h} = e\tau_{e,h}/|m^*_{e,h}|$. Henceforth we set $m^*_e \approx 0.1\,m_0$, $m^*_h \approx -0.5\,m_0$ in our calculations and consider a Γ-valley semiconductor with $d_{v,c} = 1$.

## Results

We plot the sheet conductivity with respect to voltage under different regimes in Fig. 3, using the parameters described previously where not explicitly stated. We see that under all regimes considered here, the MR is maximized at charge neutrality, with a small window of approximately 1-10 mV for highly performant magnetic readout. We also see that this window closes rapidly with respect to decreasing temperature, despite the increasing maximum MR. Additionally, given that the memtransistor only displays a memory functionality at charge neutrality, the sheet conductance is quite low, of order $10^{-5}\,\Omega^{-1}$. We have chosen the example intrinsic bandgaps in Fig. 3 to permit the Boltzmann statistics approximation (thus $E_g^P > k_B T$), while trying to maximize the on-state resistance, though this should be generalized for MQW films which may fully close the gap in the P configuration, and permit high transconductance in a MQW-FET device.

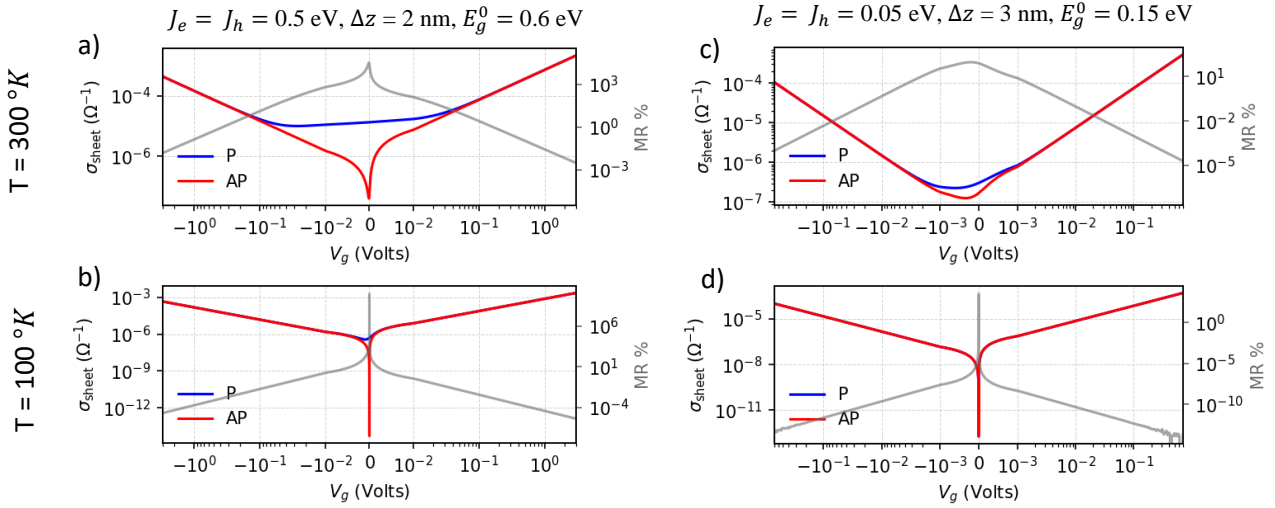

**Fig 3.** The predicted sheet conductivity and magnetoresistance of MQW thin films are shown, plotting Eq. 22 and Eq. 23 as a function of gate voltage. Blue lines demonstrate the low-bandgap configuration in the P state, while the red line shows the AP conductivity. MR is depicted in grey. a,b) show the $J_e = J_h = 0.5$ eV, $\Delta z = 2$ nm, $E_g^0 = 0.6$ eV case, while c,d) show a pessimistic $J_e = J_h = 0.05$ eV, $\Delta z = 3$ nm, $E_g^0 = 0.15$ eV case. a,c) show the device at 300 K, and b,d) show the device at 100 K.

At charge neutrality, we can discuss the analytically-derived MRs. For a QW thickness of 2 nm and a highly optimistic proximity $J_e = J_h$ of 0.5 eV, we see the possibility for a confinement-driven MR upwards of 25,000% at room-temperature, at least insofar as the mean-field exchange splitting model is faithful at a fixed $J$ at one temperature. In a system with poor proximity exchange $J_e = J_h$ of 0.05 eV, and a 3 nm QW thickness, an estimated MR of 94% is still comparable to many MTJs fabricated in academic laboratories, and should be readily detected at room temperature. We may also imagine a pristine 2-monolayer film of 1 nm (i.e. faithful to the pseudospin model in this work), with a tractable $J_e = J_h$ of 0.1 eV which predicts an MR of 1550%. This ultrathin limit, or possibly the use of 2D magnetic semiconductors as a QW, may be an attractive option as opposed to chasing an enormous magnetic exchange splitting.

## Discussion

We have shown that FMI/QW/FMI sandwiches may in theory permit enormous MRs, by using a simple 4-band model, and have outlined the essential device physics. We note that more faithful models may consider the position dependence of the proximity exchange profile, as well as the spin-orbit coupling in a 2DEG, or the temperature dependence of $J$ from the $M(T) = (1 - T/T_c)^\beta$ critical power law (using the Curie temperature $T_c$ of the FMI).

One notable result is the large on-state (P) resistance from operating in the high-MR intrinsic regime, for the parameters considered. For comparison, a square MTJ with a low resistance-area (RA) product of $5 \: \Omega \cdot \mu m^2$ [12], and side length of 40 nm will have a resistance of 3125 $\Omega$, while a MQW-FET with the same channel footprint (in the diffusive regime, and neglecting source-drain tunneling) and $\sigma_{\text{sheet}} \approx 10^{-5} \: \Omega^{-1}$ will have a resistance $R = 100 \: k\Omega$, approximately 30 times higher than the MTJ. We may also consider intrinsic graphene at room temperature, i.e. the off-state in a graphene FET, with $\sigma_{\text{sheet}} \approx 1.5 \times 10^{-4} \: \Omega^{-1}$ [13], which is still 10 times more conductive. While this high on-state (P) resistance is broadly undesirable for low-power devices, the pursuit of a reliable memtransistor allows for novel logic circuits which permit innately lower-power operation by overcoming the memory bottleneck [14], [15]. Additionally, crossbar arrays, as employed in compute-in-memory architectures, require highly resistive memristor elements, allowing for larger analog matrices to be multiplied if the interconnect resistivity is negligible in comparison [16]. Reduction of the high P state resistance should thus be pursued through maximization of the carrier mobility (which may permit orders of magnitude in improvement in high quality quantum wells, from the $\tau$ considered) or minimization of the intrinsic gap $E_g^0$ to increase carrier concentrations at charge neutrality.

This work should motivate the search for suitable candidate materials, both for the QW and FMI, and the deposition and fabrication of related heterostructures and devices. Namely, this could benefit from QW materials which have native ferromagnetic semiconducting behaviors under certain temperatures or doping regimes [17], [18], [19]. We suggest YIG as the FMI for its known capabilities in other proximity-magnetic devices [20], [21], or possibly 2D FMIs such as CrI$_3$ for cryogenic devices with clean interfaces.

Finally, we note that the carrier concentration and bandgap modulation should readily permit optical characterization of the films, which if strong enough may be generalized to magneto-optical or magneto-plasmonic devices. Further developments of the device (spin-orbit torque write, tunnel-FET behavior, efficient skyrmion readout) will also be highly desirable to advance beyond-MTJ spintronics for advanced computing applications.

**Acknowledgements**
We thank TSMC for financial support, as well as the first-year Applied Physics graduate research assistant stipend at Northwestern University.